\documentclass[final,authoryear,3p,times]{elsarticle}
\usepackage{graphics}
\usepackage{amssymb}
\usepackage{hyperref}

\journal{Journal of Informetrics}

\begin{document}
\begin{frontmatter}

\title{Zipf's law and log-normal distributions in measures of scientific output across fields and institutions: 40 years of Slovenia's research as an example}

\author{Matja{\v z} Perc\corref{mp}\corref{supp}}
\cortext[mp]{Electronic address: \href{mailto:matjaz.perc@uni-mb.si}{matjaz.perc@uni-mb.si}; Homepage: \href{http://www.matjazperc.com/}{http://www.matjazperc.com/}}
\cortext[supp]{Supplementary tables for this paper are accessible via:  \href{http://www.matjazperc.com/sicris/stats.html}{http://www.matjazperc.com/sicris/stats.html}}
\address{Department of Physics, Faculty of Natural Sciences and Mathematics, University of Maribor,\\Koro{\v s}ka cesta 160, SI-2000 Maribor, Slovenia}

\begin{abstract}
Slovenia's Current Research Information System (SICRIS) currently hosts 86,443 publications with citation data from 8,359 researchers working on the whole plethora of social and natural sciences from 1970 till present. Using these data, we show that the citation distributions derived from individual publications have Zipfian properties in that they can be fitted by a power law $P(x) \sim x^{-\alpha}$, with $\alpha$ between $2.4$ and $3.1$ depending on the institution and field of research. Distributions of indexes that quantify the success of researchers rather than individual publications, on the other hand, cannot be associated with a power law. We find that for Egghe's g-index and Hirsch's h-index the log-normal form $P(x) \sim \exp[-a\ln x -b(\ln x)^2]$ applies best, with $a$ and $b$ depending moderately on the underlying set of researchers. In special cases, particularly for institutions with a strongly hierarchical constitution and research fields with high self-citation rates, exponential distributions can be observed as well. Both indexes yield distributions with equivalent statistical properties, which is a strong indicator for their consistency and logical connectedness. At the same time, differences in the assessment of citation histories of individual researchers strengthen their importance for properly evaluating the quality and impact of scientific output.
\end{abstract}


\end{frontmatter}

\section{Introduction}

Raking of researchers is both important as well as interesting. While importance is largely due to the determination of advancement and selection criteria that underly faculty recruitments or the awarding of research grants and funds to individuals with best indicators \citep{garfieldXcc83, adamXnat02, venturaXsciento06}, the fact that it is interesting has many more aspects worth considering. For one, researchers seem to have a keen interest for determining who is the most cited or the most connected or the most influential of them all. Certainly this in part to gratify the personal sense of achievement, but more intricately, there is a lot we don't yet understand in terms of how and why certain researchers get more attention than others, and why some cannot rise above a given level of recognition. Scientific excellence is definitely a crucial factor to consider, yet that alone cannot explain all the fascinating properties that have been revealed in recent years with regards to citation distributions \citep{eggheXbook, laherrereXepjb98, rednerXepjb98, rednerXphystoday05, radicchiXpnas08, vieiraXjinfor10}, indexes that quantify individual scientific output \citep{hirschXpnas05, eggheXscient06, eggheXjasoc08, bornmannXjasoc08, zhangXplos09, gunsXjinfor09, cabrerizoaXjinfor10}, the importance of first-movers \citep{newmanXepl09} and self-citations \citep{folwerXsciento07, schreiberXepl07, schreiberXsciento08}, or the structure of scientific collaboration networks \citep{newmanXpnas01}, to name but a few.

Empirical studies are important since they provide fuel for potential attempts at modeling and related theoretical approaches aimed towards deepening our understanding of citation practices, as well as for sharpening criteria and indexes that quantify individual scientific output. Notably, one fact stands quite solid and has been pointed out on several occasions [see \textit{e.g.} \citet{rednerXphystoday05}]. Namely that the more one paper is cited, the more likely it is it will attract further citations in the future. This phenomenon is by now known under different names. The Matthew effect \citep{mertonXsci68} is likely the oldest to describe it, but one can come across also cumulative advantage \citep{priceXsci65, priceXjamsoc76} or preferential attachment \citep{barabaXsci99}, depending on the field of research and motivation of the study. Especially linear preferential attachment models enjoy exceptional popularity in describing the growth and setup of complex networks \citep{barabaXrmp02, dorogovtsevXbook03, pastorXbook04} and have become synonymous for power-law distributions of connections that can be observed in many of them \citep{faloutsos99, sornetteXbook03, newmanXcp05, clausetXsiam09}. There is evidence suggesting that citation statistics may obey to similar rules, yet deviations from the power-law distribution maintain the reasoning open to amendments \citep{rednerXphystoday05}, especially in the sense of sublinear or near-linear preferential attachment, which is know to yield stretched exponential or log-normal forms \citep{krapivskyXprl00, dorogovtsevXepl00, dorogovtsevXprl00, krapivskyXprl01, krapivskyXpre01}.

Here we present the analysis of 40 years of Slovenia's research output across the whole of social and natural sciences in search for signs of self-organization and laws that underly many aspects of our existence. Zipf's law \citep{zipfXbook49} in particular is related to the frequent occurrence of power-law distributions, with examples ranging from the frequency of words in a given language, income rankings, population counts of cities to avalanche and forest-fire sizes \citep{newmanXcp05}.\footnote{A comprehensive list of publications devoted to the Zipf's law is accessible via: \href{http://www.nslij-genetics.org/wli/zipf/}{http://www.nslij-genetics.org/wli/zipf/} (by \href{http://www.nslij-genetics.org/wli/}{Wentian Li})} We show that the citation distributions derived from individual publications, \textit{i.e.} determined as the number of publications with a certain number of citations, are of power-law type, which indeed seems to confirm the assumption of linear preferential attachment underlying their accumulation. However, by taking into consideration not individual publications but rather individual researchers, we find that the power-law distributions give way to log-normal, and in special cases also exponential \citep{laherrereXepjb98}, distributions. Notably, both the g-index \citep{eggheXscient06} and the h-index \citep{hirschXpnas05}, as well as the total citation count per researcher, show equivalent statistical properties in terms of their distributions. This suggests that these measures share a relatively high degree of logical connectedness that cannot be distinguished on large scales. However, differences between them can be crucial for the ranking of individual researchers within specific groups or fields of research. Since log-normal forms are typically associated with random multiplicative processes, the assumption of liner preferential attachment as the main driving force behind the citation record of an individual researcher seems no longer valid. Certainly it plays a role, but the ``personality'' of a researcher brings with it additional factors that require a different interpretation. An important role seems to play the fact that all researchers more or less frequently publish papers that don't receive a lot of attention. At the same time, a researcher can gather a considerable number of citations even if s/he doesn't publish a single highly-cited paper. Altogether, these considerations, which are absent when considering individual publications as reference points, amount to an override of the power-law distribution. We also point out that, as discovered already by \citet{rednerXepjb98}, not a single function can describe the examined distributions over the whole range of values. Power laws emerge due to collective effects, synonymous to preferential attachment, which apply to well-cited publications only. Papers that are not cited frequently do not benefit from such or similar effects and are forgotten soon after their publication. Presented results thus fit well to known facts, as well as provide a cohesive overview of factors that affect the distributions of citations and other measures of scientific output.

The paper is structured as follows. In the next section we provide basic facts about Slovenia and the analyzed data set. We also review basic properties of Zipf plots, power-law and log-normal distributions, which will be called upon when presenting the main results in section 3. In the last section we summarize our findings and briefly discuss their implications for the national selection criteria currently employed by the Slovenian Research Agency.

\section{Preliminaries}

Slovenia is a small country located at the heart of Europe with a population of two million.\footnote{The official Web page of Slovenia is accessible via: \href{http://www.slovenia.si/}{http://www.slovenia.si/}} It has a very well-documented research history, which is made possible by SICRIS -- Slovenia's Current Research Information System.\footnote{The SICRIS Web page is accessible via: \href{http://sicris.izum.si/}{http://sicris.izum.si/}} At present, Slovenia has 30,630 registered researchers (including young and non-active researchers as well as laboratory personnel), of which 8,359 have at least one bibliographic unit that is indexed by the Web of Science (WoS). Currently there are 86,443 publications linked to WoS with a total of 835,970 citations that have accumulated from 1970 till present. Bibliographies of researchers are updated continuously by a group of specialized libraries that catalogue new publications as soon as they are collated, while the citation data of all bibliographic units are updated monthly via a direct link to WoS.

Since the SICRIS database is publicly available, we have retrieved full publication records by means of an automated information retrieval algorithm, allowing us to keep the statistics as up-to-date as possible. Subsequently, the bibliographic records were parsed for citation counts and other measures that are relevant for assessing the scientific output of individual researchers. Besides analyzing the data as a whole, we consider separately the University of Ljubljana (Slovenia's oldest and largest University) and the ``Jo{\v z}ef Stefan'' Institute (Slovenia's leading research Institute), as well as researchers that designated medicine or chemistry as their primary research fields. Since the tables are too big to fit here, we made them available online at \href{http://www.matjazperc.com/sicris/stats.html}{http://www.matjazperc.com/sicris/stats.html}. The Web page features tables made also for a few other institutions and fields of research, but here we focus on the representative and most interesting examples listed above. Note that the tables can be ordered according to different categories. Some trivia:\footnote{Based on publication records retrieved in January 2010.} Slovenia's most cited researcher to date is Robert Blinc, having 10,891 citations to his name. Slovenia's most cited paper, currently having 1,374 citations, is due to Latif \textit{et al.}, entitled ``Identification of the von Hippel-Lindau disease tumor suppressor gene", which appeared in Science 260, 1317-1320 (1993). The largest g-index has Uro{\v s} Seljak (92), while the largest h-index has Vito Turk (53). From the 86,443 publications indexed by WoS 22,730 are uncited, 23,206 are cited at least 10 times, 729 are cited at least 100 times, while 8 have more than 1,000 citations.

In what follows, we first examine the distributions of citations to individual papers, whereby we first construct Zipf plots of the number of citations $x_k$ versus the $k$-th ranked paper. On a double logarithmic scale a usable linear fit of the Zipf plot with slope $\gamma$ indicates a power-law distribution of citations $P(x) \sim x^{-\alpha}$, where $\alpha=1+1/\gamma$. Likewise, the cumulative distribution of citations $Q(x)$, defined as the probability that a paper has at least $x$ citations, is proportional to $x^{-\beta}$, where $\beta=\alpha-1=1/\gamma$. Note that the joint consideration of distributions and cumulative distributions, besides the fact that the later alleviates statistical fluctuations, is useful since it helps to pinpoint the presence of a power law. Namely if $P(x) \sim x^{-\alpha}$ (is a power-law with slope $\alpha$), then also $Q(x)$ will be a power-law, but with the slope $\alpha-1$ rather than $\alpha$. On the other hand, if $P(x) \sim \exp^{-x/\kappa}$ (is exponential with slope $\kappa$) then $Q(x)$ will also be exponential, but with the same exponent \citep{newmanXsiam03}. Thus, plotting $P(x)$ and $Q(x)$ on logarithmic or semi-logarithmic scales makes it easy to distinguish power-law from exponential distributions. In a similar fashion, we subsequently construct Zipf plots of the g-index $g_k$ and the h-index $h_k$ versus the $k$-th ranked researcher, as well as plot the pertaining cumulative distribution functions $Q(g)$ and $Q(h)$. Unlike for individual publications, the Zipf plots have a negative curvature on a double logarithmic scale or can be fitted by a straight line on a semi-log scale, which indicates $Q(g) \sim \exp[-a\ln g -b(\ln g)^2]$ or $Q(g) \sim \exp(-g/\kappa)$, respectively. For individual researchers we don't consider the classical distributions of the g-index $P(g)$ and the h-index $P(h)$ since the statistical fluctuations are too strong, especially for the considered subsets of the whole population. All nonlinear fits presented in this paper have been made with the Levenberg-Marquardt method \citep{pressXbook95}, and the goodness-of-fits has been tested by means of the coefficient of determination $R^2$. Since, however, this procedure can yield substantially inaccurate fits, we have also performed maximum-likelihood fitting and the $p$-value test, as advocated in the review by \citet{clausetXsiam09}.\footnote{A comprehensive set of methods for fitting power laws accompanying the review is available via: \href{http://www.santafe.edu/~aaronc/powerlaws/}{http://www.santafe.edu/~aaronc/powerlaws/}} Given that $Q(g)$ and $Q(h)$ have equivalent statistical properties, we finally plot the relative ranks (we first rank the researchers according to one indicator and subsequently the ordered set of numbers is ranked again according to a second indicator) of researchers as determined by the g-index, the h-index, and the total citation count, showing that maximal deviations of individual rankings increase with the rank number, but remain uniformly distributed with respect to the diagonal throughout the set. Absolute values of the indicators are depicted in support of this as well, in turn implying their statistical equivalence, but at the same time strengthening their importance for individual ranking within specific groups of researchers.

\section{Results}

\begin{figure}[ht]
\begin{center}
\includegraphics[width=16.5cm]{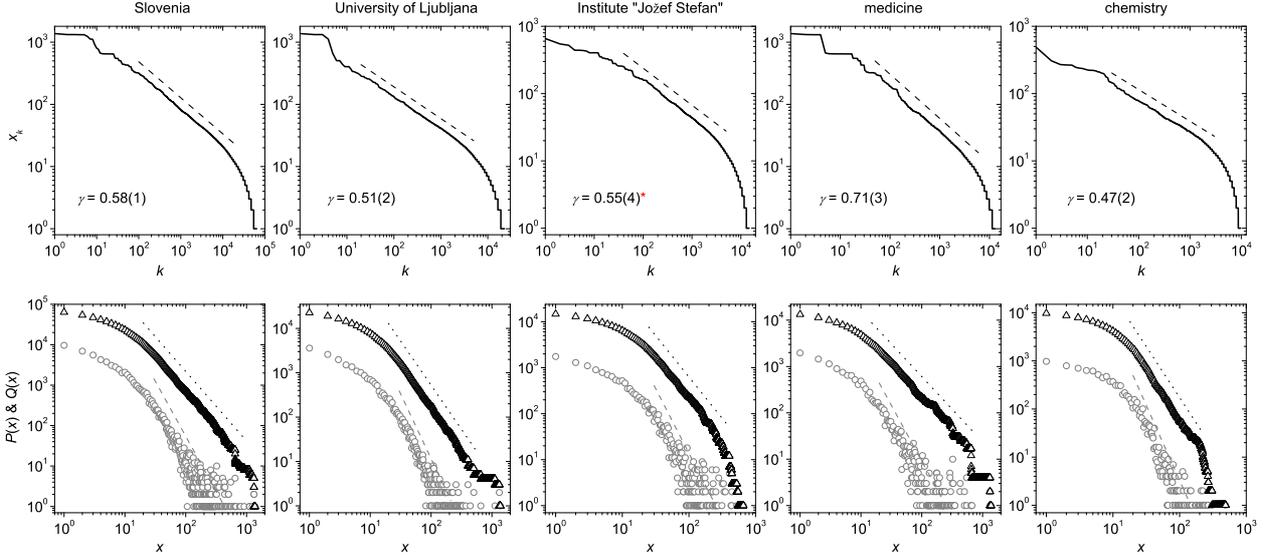}
\caption{\textit{Top row} - Zipf plots of the number of citations $x_k$ versus the $k$-th ranked paper on a double logarithmic scale. Dashed lines with slope $\gamma$ in each panel are data fits depicted for visual reference. The red star by the $\gamma$ value in the middle panel indicates that for the Institute ``Jo{\v z}ef Stefan" the fit applies to a considerably narrower region than in the other panels. \textit{Bottom row} - Citation distributions $P(x)$ (gray $\circ$) and cumulative citation distributions $Q(x)$ (black $\bigtriangleup$) obtained from the number of citations $x$ to individual publications. Dashed gray and dotted black lines with slopes $\alpha=1+1/\gamma$ and $\beta=1/\gamma$, respectively, where $\gamma$ is taken from the corresponding top panels, are depicted for visual reference. Fitting the depicted cumulative citation distributions directly yields (from left to right): $\beta=1.70(1), x_{{\rm min}}=22, R^2=0.999$; $\beta=1.92(1), x_{{\rm min}}=25, R^2=0.999$; $\beta=1.75(2), x_{{\rm min}}=26, R^2=0.996$; $\beta=1.36(1), x_{{\rm min}}=13, R^2=0.997$; $\beta=2.06(2), x_{{\rm min}}=18, R^2=0.997$, where $x_{{\rm min}}$ is the lower bound of the power-law behavior \citep{clausetXjcr07} and $R^2$ is the coefficient of determination. In the middle panel the $p$-value is lower than $0.1$, thus making $Q(x) \sim x^{-\beta}$ a questionable model for the data. Numbers in parentheses give the error on the last figure.}
\label{cites}
\end{center}
\end{figure}

We start by presenting Zipf plots of the number of citations $x_k$ versus the $k$-th ranked paper on a double logarithmic scale in the top row of Fig.~\ref{cites}. Results are presented separately for Slovenia (all 86,443 publications; 835,970 citations; 9.67 per paper), for the University of Ljubljana (subset of 30,767 publications; 263,958 citations; 8.58 per paper), for the ``Jo{\v z}ef Stefan'' Institute (subset of 17,425 publications; 230,700 citations; 13.24 per paper), as well as for medicine (subset of 19,220 publications; 195,119 citations; 10.15 per paper) and chemistry (subset of 11,370 publications; 126,055 citations; 11.09 per paper) as two representative fields of research. Apart from deviations at low and high values of $k$, it is possible to fit a straight line reasonably well to the plots with the least-squares fit yielding the exponents $\gamma$ as depicted in the corresponding panels. Notably, for the ``Jo{\v z}ef Stefan'' Institute the Zipf plot has a slight negative radius across the whole span of $k$, thus making the appropriateness of the linear fit debatable (marked with the red star). In any case, the ``Jo{\v z}ef Stefan'' Institute is special in that its publications have a comparatively high average of citations per paper (13.24 compared to the national average of 8.58), and that in the past it had a rather strict hierarchical constitution. Depending on the considered set of publications, $\gamma$ ranges from $0.47-0.71$, which theoretically corresponds to power-law distributions $P(x) \sim x^{-\alpha}$ with $\alpha$ between $2.41-3.13$, or equivalently to cumulative power-law distributions $Q(x) \sim x^{-\beta}$ with $\beta$ between $1.41-2.13$.

The bottom row of Fig.~\ref{cites} features $P(x)$ (gray $\circ$) and $Q(x)$ (black $\bigtriangleup$) of the corresponding Zipf plots from the top row. It can be observed that the Zipf plots translate fairly accurately to their expected power-law cumulative distributions $Q(x) \sim x^{-\beta}$, with Levenberg-Marquardt fits of the large-$x$ values, \textit{i.e.} $x \geq x_{{\rm min}}$, delivering exponents in agreement with $\beta \approx 1/\gamma$ (see the caption of Fig.~\ref{cites} for details). Moreover, the corresponding distributions $P(x)$ also show power-law properties in that $P(x)\sim x^{-\alpha}$ on a double logarithmic scale, with $\alpha \approx \beta+1$. Altogether, these results are in good agreement with those presented earlier by \citet{rednerXepjb98}, where also the distribution of citations to individual publications that were catalogued by the Institute for Scientific Information and 20 years of publications in the Physical Review D were found to have a large-$x$ power law decay $P(x) \sim x^{-\alpha}$ with $\alpha \approx 3$. Here we show that these observations are fairly robust to variations in research fields and institutions, and can indeed be observed for a nation as a whole. Moreover, the prevalence of the Zipf law in citations to individual publications across different research fields and institutions directly implies that the mechanisms underlying this phenomenon are robust as well. The cumulative advantage \citep{priceXsci65, priceXjamsoc76} of highly cited papers thus works irrespective of particularities that can be associated with individual publications. On the other hand, it is also known that considering individual researchers as points of reference rather than individual publications can lead to rather different results. In particular, \citet{laherrereXepjb98} reported the occurrence of stretched exponentials rather than power laws when examining the distributions of citations of most cited physicists. We therefore perform a similar statistical analysis as presented in Fig.~\ref{cites} also for individual researchers.

\begin{figure}
\begin{center}
\includegraphics[width=16.5cm]{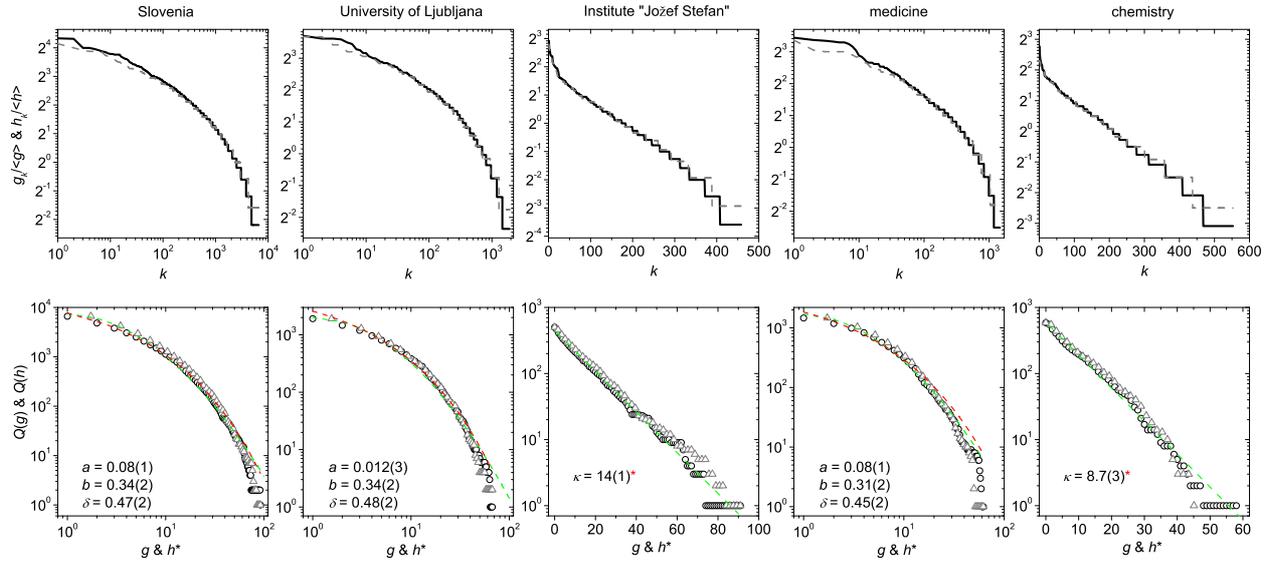}
\caption{\textit{Top row} - Zipf plots of the g-index $g_k$ (solid black) and h-index $h_k$ (dashed gray) versus the $k$-th ranked researcher on a double logarithmic or semi-log (middle and rightmost panel) scale. For comparisons, it is useful to define a scaled $k$-th ranked g-index and h-index by $\langle g \rangle$ and $\langle h \rangle$, respectively, where $\langle \cdot \rangle$ indicates average over the corresponding researcher population. \textit{Bottom row} - Cumulative g-index $Q(g)$ (black $\circ$) and h-index $Q(h)$ (gray $\bigtriangleup$) distributions obtained from the corresponding researcher population. For comparisons, the h-index on the horizontal axis was rescaled ($h \rightarrow h^*$) to fit to the interval of the g-index. Green dashed lines indicate log-normal fits of the form $Q(g) \sim \exp[-a\ln g -b(\ln g)^2]$, where the values of $a$ and $b$ are depicted in each panel. Where applicable, red dashed lines indicate stretched exponential fits of the form $Q(g) \sim \exp(-g^\delta)$, where the values of $\delta$ are depicted in each panel. In the middle and rightmost panel, however, the distribution is not log-normal but exponential, such that $Q(g) \sim \exp(-g/\kappa)$, where $\kappa \approx 14(1)$ and $\kappa \approx 8.7(3)$, respectively. Numbers in parentheses give the error on the last figure. The goodness-of-fit as determined via $R^2$ is beyond $0.99$ in all cases, except for the stretched exponential fits where it equals $0.97$.}
\label{indexes}
\end{center}
\end{figure}

Zipf plots of the g-index $g_k$ (solid black) and h-index $h_k$ (dashed gray) versus the $k$-th ranked researcher on a double logarithmic or semi-log scale (depending on the considered set of researchers) are presented in the top panel of Fig.~\ref{indexes}. As above, results are presented separately for Slovenia (all 8,359 researchers), for the University of Ljubljana (subset of 2,377 researchers), for the ``Jo{\v z}ef Stefan'' Institute (subset of 501 researchers), as well as for medicine (subset of 1,684 researchers) and chemistry (subset of 588 researchers). By comparing these results to those presented in the top row of Fig.~\ref{cites}, it becomes clear that in case of individual researchers power laws are no longer possible to advocate. The curves either have a negative radius across the whole set of $g_k$ and $h_k$ values, or can be fitted by a straight line on a semi-log scale (middle and rightmost panel). Furthermore, it is remarkable to observe that the g-index and the h-index (as well as the total citation count; not shown) have equivalent statistical properties in terms of their Zipf plots as well as the corresponding cumulative distributions $Q(g)$ and $Q(h)$, which are shown in the bottom row of Fig.~\ref{indexes}. We find that the best fits to the cumulative distributions are obtained either by means of a log-normal $Q(g) \sim \exp[-a\ln g -b(\ln g)^2]$ or an exponential $Q(g) \sim \exp(-g/\kappa)$ function, where the values of $a$, $b$ and $\kappa$ (where applicable) are depicted in the corresponding panels. Notably, the departure from the log-normal to the exponential distribution can be observed for the ``Jo{\v z}ef Stefan'' Institute (middle panel) and for the research field of chemistry (rightmost panel). Although it is difficult to pinpoint exactly why this happens, some clues can be gathered from the self-citation rates. The national average is 0.19, meaning that 160,725 from the total of 835,970 citations are self-citations. The University of Ljubljana has 0.22 (59,988 out of 263,958), the ``Jo{\v z}ef Stefan'' Institute has 0.20 (46,940 out of 230,700), medicine has 0.13 (26,284 out of 199,947) while chemistry has 0.31 (38,659 out of 124705). From these values it can be concluded that fields of research with a relatively high self-citation rate, such as chemistry in our case, are more likely to yield exponential distributions of scientific output related to individual researchers. Regarding the ``Jo{\v z}ef Stefan'' Institute, which also features an exponential $Q(g)$, we have already noted its past rather strict hierarchical constitution, which may have adversely affected the ranking of subordinate individuals (or promoted the ranking of superior individuals). It is worth noting that the log-normal form applied in the bottom row of Fig.~\ref{indexes} (green dashed line) can in our case be replaced fairly well also by a stretched exponential $Q(g) \sim \exp(-g^\delta)$ (red dashed line), which was reported by \citet{laherrereXepjb98}, thus making our results essentially in agreement with earlier works and extending their validity beyond specific fields of research as well as institutions. Lastly regarding the results presented in Fig.~\ref{indexes}, it is interesting to note that log-normal distributions were reported recently also by \citet{rednerXphystoday05} for the citation data of 110 years of the Physical Review. Although there individual papers were taken as points of reference, and one could therefore expect the prevalence of power-law distributions in accordance with earlier works \citep{rednerXepjb98} and our Fig.~\ref{cites}, the fact that only internal citations (\textit{i.e.} citations from Physical Review articles to other Physical Review articles) were considered might have been a factor contributing to the deviation.

\begin{figure}
\begin{center}
\includegraphics[width=11.46cm]{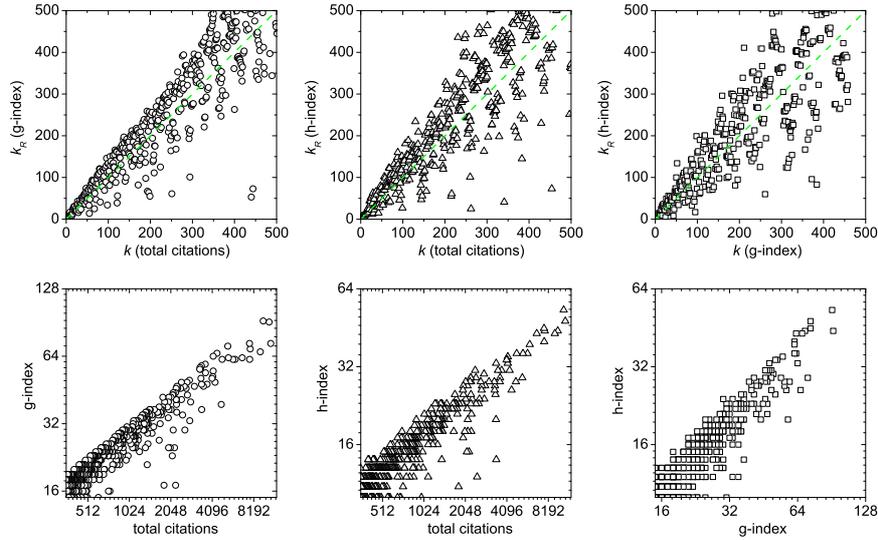}
\caption{\textit{Top row} - Comparison of researcher rankings based on different indicators of their scientific output. Researcher are first ranked according to one indicator. Subsequently, the obtained ordered set $k$ is reordered according to the ranking of researchers based on a second indicator, thus yielding the relative rank $k_R$. Plotting $k$ versus $k_R$ shows to what extend the ranking via the two considered indicators differs. If all point would fall on the diagonal (depicted dashed green for visual reference), this would imply that the two indicators yield an identical ranking of the considered set of researchers. Compared pairs of indicator are (from left to right): total number of citations versus the g-index, total number of citations versus the h-index, and g-index versus the h-index. \textit{Bottom row} - Comparisons of absolute values of the indicators, corresponding to the pairs considered in the top panels. A double logarithmic scale is used because of the substantially different maximal values of the compared indicators. Note also that the top-seeded researchers in this representation are positioned top right rather than bottom left. In all the panels top $500$ researchers are displayed.}
\label{ranking}
\end{center}
\end{figure}

With respect to the statistical equality of distributions of the g-index and the h-index (as well as the total citation count; not shown) it is instructive to examine relative rankings of pairs of different indicators. First ordering the researchers by rank according to their total citation count, \textit{i.e.} their total number of citations, and then ranking again the ordered set of numbers according to the g-index, yields how (and in which direction) the ranking of an individual differs when evaluated via the total citation count or via the g-index. This can be made for different combinations of scientific output indicators, as presented in the top row of Fig.~\ref{ranking} for the top 500 researchers of Slovenia. It can be observed that differences in ranking are indeed present, but they seem equally probable in both directions for any given $k$ -- it is not as if a given indicator would systematically downgrade only those with low $k$, for example. It is also interesting to note that the deviations from the diagonal become larger with increasing $k$, which indicates that lower-ranking researchers are more likely to be rated differently by different measures, while high-ranking researchers will remain top-seeded irrespective of which indicator is used. Importantly, however, this observation is not entirely surprising because, as we move towards the lower rankings, more and more researchers will have the same indicator so that small absolute changes of the indicator are more likely to lead to large changes in the rank. We therefore show in the bottom row of Fig.~\ref{ranking} the pertaining comparisons of absolute values of the different indicators for the top 500 researchers, which however, confirm to a large extend that the ranking via different indicators is more likely to deviate for lower-ranking than for the top-seeded researchers. Given the definitions of the g-index \citep{eggheXscient06} and the h-index \citep{hirschXpnas05}, as well as their relatedness to the total citation count, these results are not surprising and confirm the consistency and logical connectedness of these measures. At the same time, they provide some justification as to why the distributions of the g-index and the h-index are practically equivalent (see Fig.~\ref{indexes}), but also point out the fact that the properties of citation records of each individual are crucial for its ranking within a given group. Different indexes and measures of scientific output \citep{hirschXpnas07, iglesiasXscient07, jinXjam07, sidiropoulosXscient07, rousseauXjam08, barilanXinfor08} are therefore extremely useful and indeed much needed to properly evaluate the quality and impact of individual researchers.

\section{Summary}

In sum, we have shown that the distributions of citations per publication for different institutions and research fields, as well as Slovenia as a whole, have Zipfian properties in that they can be fitted fairly accurately by a power law. On the other hand, taking into account individual researchers rather than publications, we have shown that the cumulative distributions of Egghe's g-index and Hirsch's h-index are consistent with a log-normal, or in case of research fields with high self-citation rates or organizations with a special constitution, an exponential form. Interestingly, the distributions of the two indexes are statistically equivalent, thus implying their consistency and logical connectedness, but at the same time also strengthening their importance for properly assessing the scientific output of individual researchers. As a cautionary note with respect to the national selection criteria currently employed by the Slovenian Research Agency (ARRS\footnote{The ARRS Web page is accessible via: \href{http://www.arrs.gov.si/}{http://www.arrs.gov.si/}}), we note that a favorable bias in ranking emerges due to not taking into account the number of co-authors when evaluating the citation data of individual researchers \citep{wanXcnet07, schreiberXjinform08, schreiberXnjp08, eggheXjama08}. Consequently, researchers that are members of collaboration networks involved in Particle Physics research (\textit{e.g.} DELPHI, Belle or HERA-B) dominate the rankings. We hope the study will be useful for deriving theoretical models \citep{eggheXjinform09} explaining the emergence of empirically observed distributions and for drawing further attention to this interesting topic.

\section*{Acknowledgments}
Matja{\v z} Perc thanks Matej Horvat from Hermes SofLab for illuminating lessons on socket programming and automated information retrieval from the Internet. Financial support from the Slovenian Research Agency (grant Z1-2032) is gratefully acknowledged as well.


\section*{References}
\begin{thebibliography}{99}
\expandafter\ifx\csname natexlab\endcsname\relax\def\natexlab#1{#1}\fi
\expandafter\ifx\csname url\endcsname\relax
  \def\url#1{\texttt{#1}}\fi
\expandafter\ifx\csname urlprefix\endcsname\relax\def\urlprefix{URL }\fi

\bibitem[{Adam(2002)}]{adamXnat02}
Adam, D., 2002. The counting house. Nature 415, 726--729.

\bibitem[{Albert and Barab\'asi(2002)}]{barabaXrmp02}
Albert, R., Barab\'asi, A.~L., 2002. Statistical mechanics of complex networks.
  Rev. Mod. Phys. 74, 47--–97.

\bibitem[{Bar-Ilan(2008)}]{barilanXinfor08}
Bar-Ilan, J., 2008. Informetrics at the beginning of the 21st century - A
  review. J. Informetrics 2, 1--52.

\bibitem[{Barab\'asi and Albert(1999)}]{barabaXsci99}
Barab\'asi, A.~L., Albert, R., 1999. Emergence of scaling in random networks.
  Science 286, 509--–512.

\bibitem[{Bornmann et~al.(2008)Bornmann, Mutz, and Daniel}]{bornmannXjasoc08}
Bornmann, L., Mutz, R., Daniel, H.-D., 2008. Are there better indices for
  evaluation purposes than the h index? A comparison of nine different variants
  of the h index using data from biomedicine. J. Amer. Soc. Inform. Sci. 59,
  830--837.

\bibitem[{Cabrerizoa et~al.(2010)Cabrerizoa, Alonso, Herrera-Viedma, and
  Herrera}]{cabrerizoaXjinfor10}
Cabrerizoa, F.~J., Alonso, S., Herrera-Viedma, E., Herrera, F., 2010.
  q$^2$-index: Quantitative and qualitative evaluation based on the number and
  impact of papers in the hirsch core. J. Informetrics 4, 23--28.

\bibitem[{Clauset et~al.(2007)Clauset, Young, and Gleditsch}]{clausetXjcr07}
Clauset, A., Young, M., Gleditsch, K.~S., 2007. On the Frequency of Severe Terrorist Attacks. Journal of Conflict Resolution 51, 58--87.

\bibitem[{Clauset et~al.(2009)Clauset, Shalizi, and Newman}]{clausetXsiam09}
Clauset, A., Shalizi, C.~R., Newman, M. E.~J., 2009. Power-law distributions in
  empirical data. SIAM Review 51, 661--703.

\bibitem[{de~Solla~Price(1965)}]{priceXsci65}
de~Solla~Price, D.~J., 1965. Networks of scientific papers. Science 149,
  510--515.

\bibitem[{de~Solla~Price(1976)}]{priceXjamsoc76}
de~Solla~Price, D.~J., 1976. A general theory of bibliometric and other
  cumulative advantage processes. J. Amer. Soc. Inform. Sci. 27, 292--306.

\bibitem[{Dorogovtsev and Mendes(2000)}]{dorogovtsevXepl00}
Dorogovtsev, S.~N., Mendes, J. F.~F., 2000. Scaling behaviour of developing and decaying networks. Europhys. Lett. 52, 33--39.

\bibitem[{Dorogovtsev et~al.(2000)}]{dorogovtsevXprl00}
Dorogovtsev, S.~N., Mendes, J. F.~F., Samukhin, A.~N., 2000. Structure of Growing Networks With Preferential Linking. Phys. Rev. Lett. 85, 4633--4636.

\bibitem[{Dorogovtsev and Mendes(2003)}]{dorogovtsevXbook03}
Dorogovtsev, S.~N., Mendes, J. F.~F., 2003. Evolution of Networks: From
  Biological Nets to the Internet and WWW. Oxford University Press, Oxford.

\bibitem[{Egghe(2006)}]{eggheXscient06}
Egghe, L., 2006. Theory and practise of the g-index. Scientometrics 69,
  131--152.

\bibitem[{Egghe(2008{\natexlab{a}})}]{eggheXjasoc08}
Egghe, L., 2008{\natexlab{a}}. The influence of transformations on the h-index
  and the g-index. J. Amer. Soc. Inform. Sci. 59, 1304--1312.

\bibitem[{Egghe(2008{\natexlab{b}})}]{eggheXjama08}
Egghe, L., 2008{\natexlab{b}}. Mathematical theory of the h- and the g-index in
  case of fractional counting of authorship. J. Amer. Soc. Inform. Sci. 59,
  1608--–1616.

\bibitem[{Egghe(2009)}]{eggheXjinform09}
Egghe, L., 2009. Mathematical derivation of the impact factor distribution. J.
  Informetrics 4, 290--295.

\bibitem[{Egghe and Rousseau(1990)}]{eggheXbook}
Egghe, L., Rousseau, R., 1990. Introduction to Informetrics: Quantitative
  Methods in Library, Documentation and Information Science. Elsevier,
  Amsterdam.

\bibitem[{Faloutsos et~al.(1999)Faloutsos, Faloutsos, and
  Faloutsos}]{faloutsos99}
Faloutsos, M., Faloutsos, P., Faloutsos, C., 1999. On power-law relationships
  of the internet topology. In: SIGCOMM '99: Proceedings of the conference on
  Applications, technologies, architectures, and protocols for computer
  communication. ACM, New York, NY, USA, pp. 251--262.

\bibitem[{Fowler and Aksnes(2007)}]{folwerXsciento07}
Fowler, J.~H., Aksnes, D.~W., 2007. Does self-citation pay? Scientometrics 72,
  427--–437.

\bibitem[{Garfield(1983)}]{garfieldXcc83}
Garfield, E., 1983. How to use citation analysis for faculty evaluations, and
  when is it relevant? Current Contents 45, 5--146.

\bibitem[{Guns and Rousseau(2009)}]{gunsXjinfor09}
Guns, R., Rousseau, R., 2009. Real and rational variants of the h-index and the
  g-index. J. Informetrics 3, 64--71.

\bibitem[{Hirsch(2005)}]{hirschXpnas05}
Hirsch, J.~E., 2005. An index to quantify an individual's scientific research
  output. Proc. Natl. Acad. Sci. USA 104, 16569--–16572.

\bibitem[{Hirsch(2007)}]{hirschXpnas07}
Hirsch, J.~E., 2007. Does the h index have predictive power? Proc. Natl. Acad.
  Sci. USA 102, 19193--19198.

\bibitem[{Iglesias and Pecharroman(2007)}]{iglesiasXscient07}
Iglesias, J.~E., Pecharroman, C., 2007. Scaling the h-index for different
  scientific ISI fields. Scientometrics 73, 303--320.

\bibitem[{Jin et~al.(2007)Jin, Liang, Rousseau, and Egghe}]{jinXjam07}
Jin, B.~H., Liang, L.~M., Rousseau, R., Egghe, L., 2007. The R- and AR-indices:
  Complementing the h-index. Chin. Sci. Bull. 52, 855--–863.

\bibitem[{Krapivsky et~al.(2000)}]{krapivskyXprl00}
Krapivsky, P.~L., Redner, S., Leyvraz, F., 2000. Connectivity of Growing Random Networks. Phys. Rev. Lett. 85, 4629--–4632.

\bibitem[{Krapivsky et~al.(2001)}]{krapivskyXprl01}
Krapivsky, P.~L., Rodgers, G.~J., Redner, S., 2001. Degree Distributions of Growing Random Networks. Phys. Rev. Lett. 86, 5401--–5404.

\bibitem[{Krapivsky and Redner(2001)}]{krapivskyXpre01}
Krapivsky, P.~L., Redner, S., 2001. Organization of growing random networks. Phys. Rev. E 63, 066123.

\bibitem[{Laherrere and Sornette(1998)}]{laherrereXepjb98}
Laherrere, J., Sornette, D., 1998. Stretched exponential distributions in
  nature and economy: ``fat tails'' with characteristic scales. Eur. Phys. J. B
  2, 525--539.

\bibitem[{Merton(1968)}]{mertonXsci68}
Merton, R.~K., 1968. The Matthew effect in science. Science 159, 56--–63.

\bibitem[{Newman(2001)}]{newmanXpnas01}
Newman, M. E.~J., 2001. The structure of scientific collaboration networks.
  Proc. Natl. Acad. Sci. USA 98, 404--–409.

\bibitem[{Newman(2003)}]{newmanXsiam03}
Newman, M. E.~J., 2003. The structure and function of complex networks. SIAM
  Review 45, 167--–256.

\bibitem[{Newman(2005)}]{newmanXcp05}
Newman, M. E.~J., 2005. Power laws, Pareto distributions and Zipf's law.
  Contemporary Physics 46, 323--351.

\bibitem[{Newman(2009)}]{newmanXepl09}
Newman, M. E.~J., 2009. The first-mover advantage in scientific publication.
  EPL 86, 68001.

\bibitem[{Pastor-Satorras and Vespignani(2004)}]{pastorXbook04}
Pastor-Satorras, R., Vespignani, A., 2004. Evolution and Structure of the
  Internet: A Statistical Physics Approach. Cambridge University Press,
  Cambridge.

\bibitem[{Press et~al.(1995)}]{pressXbook95}
Press, W.~H., Teukolsky, S.~A., Vetterling, W.~T., Flannery, B.~P., 1995. Numerical Recipes in C. Cambridge University Press, Cambridge.

\bibitem[{Radicchi et~al.(2008)Radicchi, Fortunato, and
  Castellano}]{radicchiXpnas08}
Radicchi, F., Fortunato, S., Castellano, C., 2008. Universality of citation
  distributions: Toward an objective measure of scientific impact. Proc. Natl.
  Acad. Sci. USA 105, 17268--17272.

\bibitem[{Redner(1998)}]{rednerXepjb98}
Redner, S., 1998. How popular is your paper? An empirical study of the citation
  distribution. Eur. Phys. J. B 4, 131--134.

\bibitem[{Redner(2005)}]{rednerXphystoday05}
Redner, S., 2005. Citation Statistics from 110 Years of Physical Review.
  Physics Today 58, 49--54.

\bibitem[{Rousseau and Ye(2008)}]{rousseauXjam08}
Rousseau, R., Ye, F.~Y., 2008. A proposal for a dynamic h-type index. J. Amer.
  Soc. Inform. Sci. 59, 1853--1855.

\bibitem[{Schreiber(2007)}]{schreiberXepl07}
Schreiber, M., 2007. Self-citation corrections for the hirsch index. EPL 78,
  30002.

\bibitem[{Schreiber(2008{\natexlab{a}})}]{schreiberXsciento08}
Schreiber, M., 2008{\natexlab{a}}. The influence of self-citation corrections
  on egghe's g index. Scientometrics 76, 187--–200.

\bibitem[{Schreiber(2008{\natexlab{b}})}]{schreiberXjinform08}
Schreiber, M., 2008{\natexlab{b}}. A modification of the h-index: The ${\rm
  h_{m}}$-index accounts for multi-authored manuscripts. J. Informetrics 2,
  211--216.

\bibitem[{Schreiber(2008{\natexlab{c}})}]{schreiberXnjp08}
Schreiber, M., 2008{\natexlab{c}}. To share the fame in a fair way, ${\rm
  h_{m}}$ modifies h for multi-authored manuscripts. New J. Phys. 10, 040201.

\bibitem[{Sidiropoulos et~al.(2007)Sidiropoulos, Katsaros, and
  Manolopoulos}]{sidiropoulosXscient07}
Sidiropoulos, A., Katsaros, D., Manolopoulos, Y., 2007. Generalized hirsch
  h-index for disclosing latent facts in citation networks. Scientometrics 72,
  253--280.

\bibitem[{Sornette(2003)}]{sornetteXbook03}
Sornette, D., 2003. Critical Phenomena in Natural Sciences, chapter 14.
  Springer, Heidelberg.

\bibitem[{Ventura and Mombr\'u(2006)}]{venturaXsciento06}
Ventura, O., Mombr\'u, A.~W., 2006. Use of bibliometric information to assist
  research policy making. A comparison of publication and citation profiles of
  full and associate professors at a school of chemistry in uruguay.
  Scientometrics 69, 287--313.

\bibitem[{Vieira and Gomes(2010)}]{vieiraXjinfor10}
Vieira, E.~S., Gomes, J. A. N.~F., 2010. Citations to scientific articles: Its
  distribution and dependence on the article features. J. Informetrics 4,
  1--13.

\bibitem[{Wan et~al.(2007)Wan, Hua, and Rousseau}]{wanXcnet07}
Wan, J., Hua, P., Rousseau, R., 2007. The pure h-index: calculating an author's
  h-index by taking co-authors into account. Collnet Journal of Scientometrics
  and Information Managemen 1, 1--5.

\bibitem[{Zhang(2009)}]{zhangXplos09}
Zhang, C.-T., 2009. The e-index, complementing the h-index for excess
  citations. PLoS ONE 4, e5429.

\bibitem[{Zipf(1949)}]{zipfXbook49}
Zipf, G.~K., 1949. Human Behavior and the Principle of Least-Effort.
  Addison-Wesley, Reading MA.

\end{thebibliography}
\end{document}